\documentclass[letter,traditabstract]{aa}  

\usepackage{graphicx}
\usepackage{txfonts}

\begin{document} 

\title{On the Robustness of Bi-Stability Jump Predictions}
\subtitle{}

\author{{Jorick S. Vink\inst{\ref{AOP}}}
    \and 
       {Gautham N. Sabhahit\inst{\ref{AOP}}}
    \and
      {Andreas\,A.\,C. Sander\inst{\ref{ARI}}}
    }

\institute{
   {Armagh Observatory and Planetarium, College Hill, Armagh BT61 9DG, N. Ireland\label{AOP}}
   \and
   {Zentrum f{\"u}r Astronomie der Universit{\"a}t Heidelberg, Astronomisches Rechen-Institut, M{\"o}nchhofstr. 12-14, 69120 Heidelberg, Germany\\\label{ARI}}
   \email{jorick.vink@armagh.ac.uk}}

\date{}

\abstract{The bi-stability jump is a long-standing theoretical prediction of
radiatively driven wind theory, associated with Fe\,\textsc{iv}/Fe\,\textsc{iii}
recombination around $T_{\rm eff}\simeq 21$--25\,kK. While most theoretical
approaches predict a strong increase in mass-loss rates across the bi-stability jump, most 
empirical mass-loss studies of OB supergiants have not revealed
the expected signature.
We computed new hydrodynamically consistent PoWR models at low and intermediate Eddington
parameters ($\Gamma_{\rm e}\simeq 0.2-0.3$) to test whether the bi-stability jump persists in
the canonical B-supergiant regime.
The PoWR models presented here predict a robust bi-stability jump, with an increase in mass-loss rate
by more than an order of magnitude and a simultaneous drop in terminal
wind velocity in line with Monte Carlo models and other co-moving frame calculations. The jump coincides with a transition in the dominant line
driver from Fe\,\textsc{iv} to Fe\,\textsc{iii}. The presence of the bi-stability jump is
not restricted to high-$\Gamma_{\rm e}$ objects and remains present for
models well below the LBV/hypergiant regime.
The persistence of the bi-stability jump in hydrodynamically consistent models at lower
$\Gamma_{\rm e}$ supports the interpretation of the bi-stability jump as a
temperature-driven ionisation effect that operates once a stationary
line-driven wind solution exists.
The continuing
discrepancy between predictions  
and empirical
population studies motivates further code-comparison work and controlled
observational tests using individual objects such as LBVs.}


\maketitle

\section{Introduction}

The bi-stability (BS) jump\footnote{There is also a second bi-stability jump at an
effective temperature of roughly 10\,kK, but this is not the focus of
this Letter.}is a theoretically predicted, rapid change in
radiatively driven wind properties around
$T_{\rm eff}\!\simeq\!21$--25\,kK,
associated with a transition from a fast, relatively tenuous wind to a
slower, denser outflow. \citet{vink99,vink00} showed that the bi-stability jump is
fundamentally a temperature-driven ionisation effect, caused by the
recombination of the dominant line driver iron from Fe\,\textsc{iv} to
Fe\,\textsc{iii}, and is not intrinsically tied to proximity to the
Eddington limit, as originally suggested \citep{PP90}.

Mass-loss rates of OB supergiants are a key uncertainty in massive-star evolution, influencing envelope stripping, the formation of Wolf-Rayet stars, and the masses of black holes.
Understanding whether the bi-stability jump operates in nature is important well beyond
the details of OB-star wind theory. It is central to interpretations of
S~Doradus variability in luminous blue variables (LBVs)
\citep{Grassitelli21}, is relevant for the suggestion that LBVs are
direct supernova progenitors \citep{KV06,Trundle08}, and has been invoked
for the formation of disks around Be and B[e] stars and aspherical winds
\citep{LP91,Pel00,Cure05}.
Yet, despite more than two decades of effort, empirical evidence for a
clear bi-stability jump signature in samples of OB supergiants remains elusive
\citep{vink00,TL05,Crow06,MP08}.

In particular, \citet{Verhamme24} report a continuous decrease of
$\dot{M}$ toward lower $T_{\rm eff}$. 
In contrast, most theoretical approaches predict a reversal in $\dot{M}$ across the
Fe\,\textsc{iv}/Fe\,\textsc{iii} transition. These include both global
Monte Carlo \citep{vink99} and work-ratio based CMFGEN \citep{petrov16} modelling, as well as more locally consistent approaches from Monte Carlo calculations \citep{vink18bsj}, METUJE
models \citep{krticka21}, and hydrodynamically consistent PoWR models that utilise CMF radiative transfer \citep{Sabh26,Bernini26}. 
By contrast,
the FASTWIND-based results of \citet{bjork23} show a more continuous
behaviour without a reversal across the transition.

The goal of this Letter is to assess the robustness of bi-stability jump
predictions in hydrodynamically consistent PoWR models at low
$\Gamma_{\rm e}$, and to clarify why most theoretical predictions and empirical mass-loss constraints appear to disagree.
The current theoretical status is that most models predict a
bi-stability jump, while some approaches predict a smoother
behaviour without a clear reversal. Figure~1 of \citet{Verhamme24} is particularly instructive in this
respect. When the effective temperature decreases within the O-star
regime, all approaches predict a decline in $\dot{M}$, which
\citet{vink99} attributed to a growing mismatch between the radiative
flux distribution and the line opacity. However, once the temperature
crosses the threshold where Fe\,\textsc{iii} driving takes over from
Fe\,\textsc{iv}, the predicted mass-loss behaviour reverses in the majority of models. 

This discrepancy is sometimes attributed to the proximity of LBVs to the
Eddington limit, but this interpretation is not supported by the
original Monte Carlo results, which show the bi-stability jump over a
wide range in $\Gamma_{\rm e}$, and dynamical Monte Carlo models indicated that the relative strength
of the BS jump increased towards {\it lower} Eddington parameters
\citep{vink18bsj}.
Finally, it is essential to distinguish
between the bi-stability jump in individual objects --- where stellar parameters
remain approximately constant while $T_{\rm eff}$ changes --- and
population-based tests using samples of stars, where additional
systematic differences may enter.

\subsection{Origin of the bi-stability concept}

The bi-stability jump was first uncovered in wind models of the LBV P\,Cygni by
\citet{PP90}, and its observational phenomenology was summarised by
\citet{lamers95}, who identified a change in terminal velocities
and inferred wind densities around spectral type B1. In the earliest
theoretical interpretations, the bi-stability jump was linked to the optical depth of
the Lyman continuum in objects close to the Eddington limit, motivating
the idea that a change in the optical depth of the Lyman continuum would
be responsible for the observed wind behaviour.
Using Monte Carlo radiative transfer, \citet{vink99,vink00} revised this
picture and demonstrated that the bi-stability jump is present over a wide range of
stellar parameters, including models far from the Eddington limit. The
bi-stability jump was shown to arise primarily from a temperature-dependent change in
the flux-weighted line opacity, associated with the recombination of
iron from Fe\,\textsc{iv} to Fe\,\textsc{iii} when $T_{\rm eff}$ drops
below $\sim$21--25\,kK\footnote{Using the global \cite{AL85} MC method, \cite{vink99} found the mass-loss bi-stability jump at 25 kK, while dynamically consistent modelling using the \cite{MV08} Lambert W formalism, \cite{vink18bsj} found it closer 21 kK.}. In this framework, the Eddington parameter may
modulate the onset and amplitude of the effect, but the physical origin
is an ionisation-driven redistribution of line driving rather than a
sudden switch in Lyman continuum optical depth.

\section{PoWR hydrodynamically consistent models}

It is useful to distinguish between the physics of wind launching and
the differential opacity effect responsible for the bi-stability jump.
Hydrodynamically consistent wind calculations can in some cases struggle
to initiate a stationary outflow because of a decrease in the radiative acceleration in the wind launching region, also known as the ``source-function dip'' \citep{GH03}. While such behaviour may reflect real physical difficulties in
one-dimensional stationary models, observations clearly demonstrate that
O-type stars ubiquitously possess strong stellar winds. This suggests that
nature finds ways to overcome the launching difficulty, potentially through
multidimensional effects such as turbulent pressure or time-dependent
structure. Once a wind solution exists, however, the
Fe\,\textsc{iv}/Fe\,\textsc{iii} ionisation transition produces a
differential change in the line driving, leading to the characteristic
mass-loss and velocity changes associated with the bi-stability jump.

\citet{Sabh26} investigated the dependence of mass-loss rates on the
Eddington parameter $\Gamma_{\rm e}$ using hydrodynamically consistent
PoWR model atmospheres. In addition to kinks in the
$\dot{M}$--$\Gamma_{\rm e}$ relation, the study revealed the presence of
two distinct bi-stability jumps. Interestingly, the bi-stable behaviour
was found to be more prevalent at lower $\Gamma_{\rm e}$ than at higher
$\Gamma_{\rm e}$. The presence of the bi-stability jump at $\Gamma_{\rm e}\simeq 0.4$
raised the question of whether bi-stability would remain robust at even
lower $\Gamma_{\rm e}$.

To test this, we here extend the \citet{Sabh26} analysis to
$\Gamma_{\rm e}\simeq 0.2$ by adopting a model with $M=40\,M_\odot$ and
$\log L/L_\odot = 5.5$. We refer to \citet{Sabh26} for details of the
PoWR setup and modelling assumptions. 
Wind clumping is incorporated using the micro-clumping approach. The clumping transitions from a smooth wind at the base ($D_\mathrm{cl}=1$) to a clumping factor of $D_\mathrm{cl}=10$ in the outer wind, with the onset of clumping at an optical depth of $\tau_\mathrm{cl}=0.1$. The assumed metallicity is Z=0.02, noting that while more recent solar compositions provide lower overall metallicities, the Fe abundance that sets the mass-loss bi-stability jump should not be affected by this choice.
Throughout this Letter
we distinguish between the inner boundary temperature $T_\star$ (defined
at $\tau_{\rm Ross}=20$) and the effective temperature $T_{\rm eff}$
(defined at $\tau_{\rm Ross}=2/3$), which is typically $\sim$1--2\,kK
lower for the models considered here. 

\begin{figure}
    \includegraphics[width=0.48\textwidth]{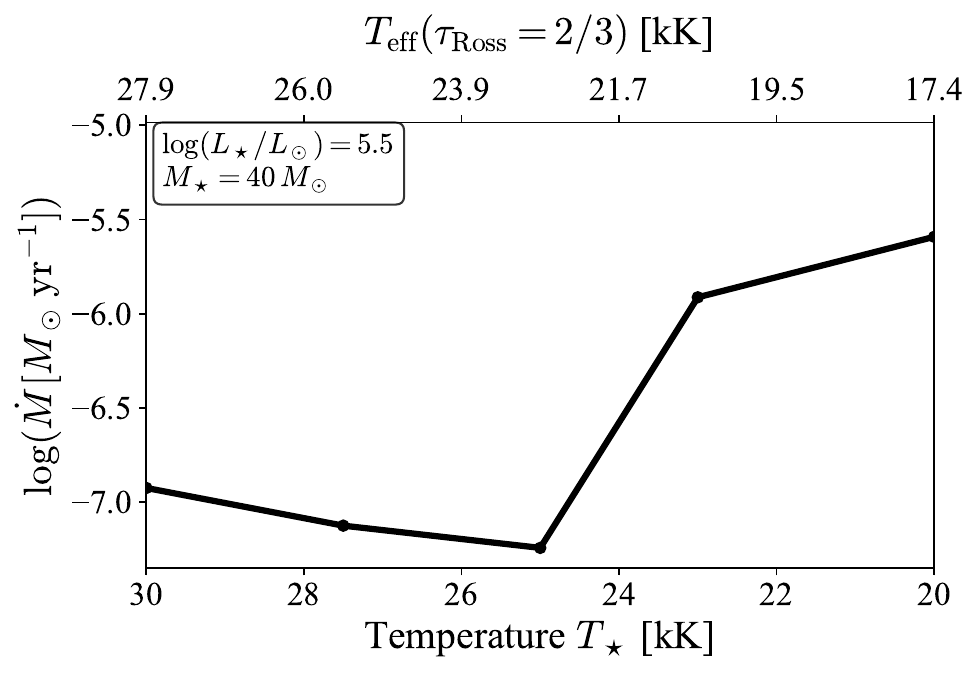}
    \caption{$\dot{M}$ versus $T_\star$, where $T_\star$ is the inner
    boundary effective temperature at Rosseland continuum optical depth
    $\tau_{\rm Ross}=20$. Model parameters: $\log L/L_\odot = 5.5$,
    $M=40\,M_\odot$ ($\Gamma_{\rm e}\simeq 0.2$), $X=0.7$, $Z=0.02$..}
    \label{mdot}
\end{figure}

\begin{figure}
    \includegraphics[width=0.48\textwidth]{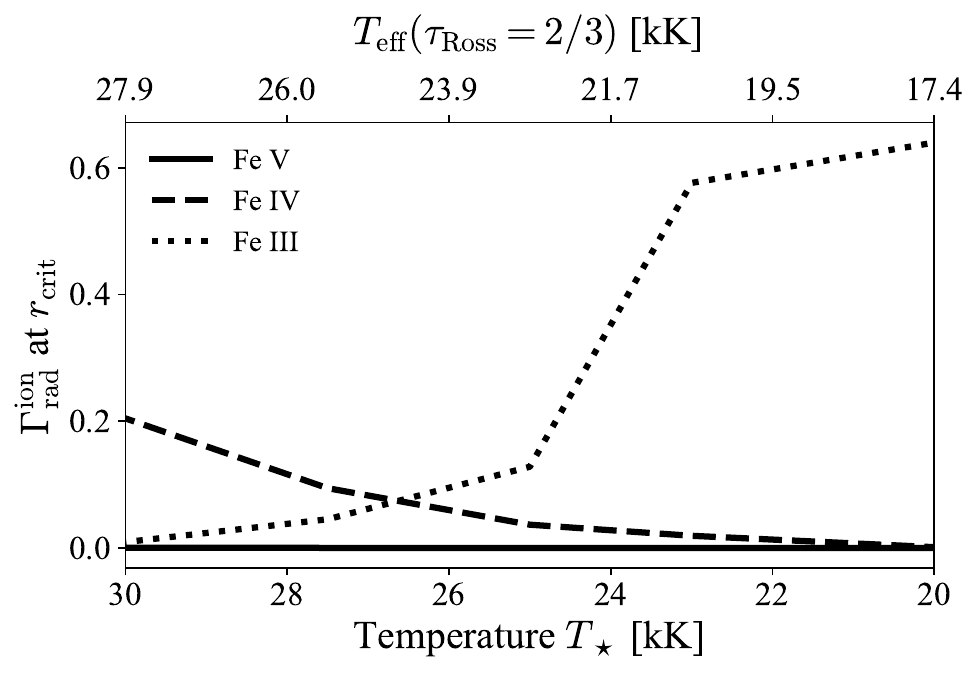}
    \caption{Radiative acceleration contributions -- normalised to gravity and expressed as Eddington parameters -- from relevant Fe wind-driving ions at the critical point, expressed as function of $T_\star$ for the same model as
    Fig.~\ref{mdot}. It illustrates the rise to Fe\,\textsc{iii} dominance across the bi-stability regime.}
    \label{ions}
\end{figure}

The mass-loss results in Fig.~\ref{mdot} show the expected jump by more
than an order of magnitude when a critical temperature is crossed, here
at an inner boundary temperature $T_\star \simeq 25$\,kK. This jump
coincides with a dramatic increase in the contribution of
Fe\,\textsc{iii} to the line driving (Fig.~\ref{ions}). The same
behaviour was previously found in Monte Carlo computations
\citep{vink99} and in CMFGEN models \citep{petrov16}.

\begin{figure}
    \centering
    \includegraphics[width=0.48\textwidth]{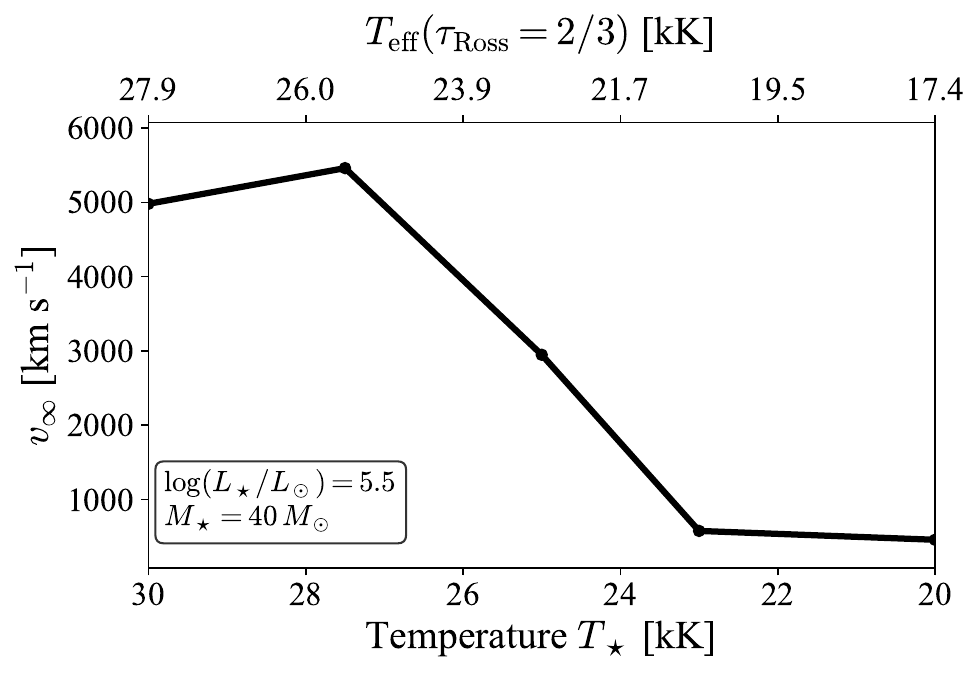}
    \caption{Terminal wind velocity $v_\infty$ versus $T_\star$ for the
    model in Fig.~\ref{mdot}.}
    \label{vinf}
\end{figure}

Figure~\ref{vinf} shows the corresponding drop in terminal wind velocity,
in line with the observational constraints by  \citet{lamers95,Crow06} and the
dynamically consistent Monte Carlo models of \citet{vink18bsj}.

The mass-loss predictions of \citet{Sabh26} have recently been adopted as
input for spectral synthesis modelling of very massive stars by
\citet{Sabh25}. In that work, a relatively high turbulent velocity\footnote{This turbulent velocity should not be confused with the microturbulence adopted in formal spectral synthesis.} was used in the hydrodynamic equation ($v_{\rm turb}=70$\,km\,s$^{-1}$), motivated by recent
multi-dimensional simulations for O-type stars \citep{Moens25}. 
However, the mass-loss jump at the bi-stability temperature is also present in
the recent PoWR modelling of \citet{Bernini26}, which was performed
using lower assumed turbulent velocities in the hydrodynamic equation than the models in
\citet{Sabh26} and likewise finds the characteristic bi-stability jumps, although those
models were calculated in a somewhat higher $\Gamma_{\rm e}$ range. 

For this reason, we performed additional sequences with more moderate
$v_{\rm turb}=21$\,km\,s$^{-1}$. The results are shown in Fig.~4 for
models with $M=30\,M_\odot$ and $M=35\,M_\odot$, corresponding to
different $\Gamma_{\rm e}$ values at fixed luminosity.
Both sequences show a clear bi-stability jump. In fact, the model with
lower $\Gamma_{\rm e}$ (in red) exhibits the larger jump, demonstrating that
the jump strength does not scale monotonically with $\Gamma_{\rm e}$.
This further supports the interpretation that the bi-stability jump is
primarily a temperature-driven ionisation effect rather than a direct
consequence of Eddington limit proximity.

\begin{figure}
\includegraphics[width=0.48\textwidth]{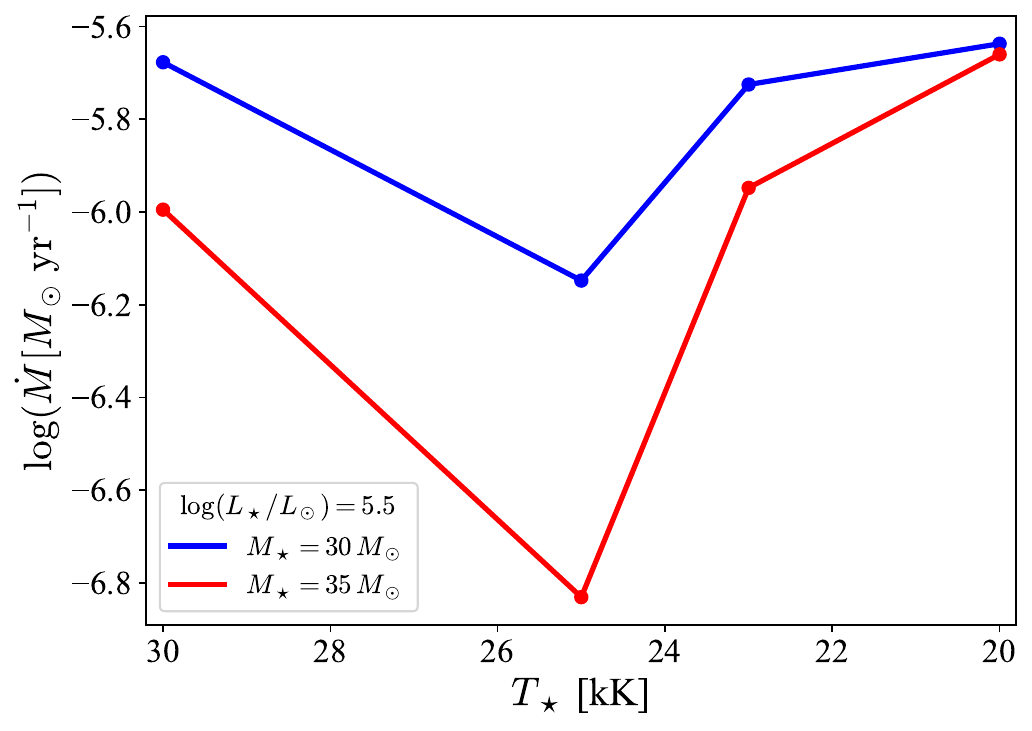}
\caption{$\dot{M}$ versus $T_\star$ for comparison models computed
with a lower adopted turbulent velocity in the hydrodynamic equation
($v_{\rm turb}=21$\,km\,s$^{-1}$). The two sequences have
$\log L/L_\odot=5.5$ and different stellar masses,
$M=30\,M_\odot$ ($\Gamma_{\rm e} = 0.28$) and $M=35\,M_\odot$ ($\Gamma_{\rm e}$ = 0.24), corresponding to different
$\Gamma_{\rm e}$. Both sequences show a bi-stability jump, with the
lower-$\Gamma_{\rm e}$ model exhibiting a larger jump.}
   \label{low-vturb}
\end{figure}

We conclude that, while multi-dimensional effects of turbulence require
substantially more rigorous investigation, current PoWR models across a
range of turbulent-pressure assumptions and $\Gamma_{\rm e}$ values
indicate that the bi-stability jump is not restricted to high-$\Gamma_{\rm e}$
objects characteristic of LBVs and hypergiants.
Instead, the
mass-loss jump remains present down to relatively low
$\Gamma_{\rm e} \sim 0.2$ values, well into the canonical B-supergiant
regime.
While proximity to the Eddington limit or the inclusion of a turbulent-pressure term can facilitate wind launching by reducing the effective gravity, the physical trigger of the bi-stability jump remains the temperature-dependent Fe\,\textsc{iv}/Fe\,\textsc{iii} ionisation balance.

\section{Why does empirical modelling not show the predicted bi-stability jump?}

The persistent absence of a clear bi-stability jump signature in empirical
mass-loss determinations for samples of OB supergiants has remained a
major puzzle for more than 25 years \citep{vink00}. We do not
attempt to resolve this discrepancy here; instead, we outline several independent
considerations that may contribute to the tension between theoretical
predictions and empirical analyses.

The bi-stability jump appears in the majority of independent modelling approaches,
including global Monte Carlo calculations, CMFGEN models, and hydrodynamically
consistent PoWR calculations. While the details of wind launching and the
treatment of line driving differ between methodologies, the presence of the
Fe\,\textsc{iv}/Fe\,\textsc{iii} opacity transition is common to these models.

\subsection{Are the empirical mass-loss determinations reliable?}

Even if the bi-stability jump is physically real, empirical modelling may not yet be
sufficiently sophisticated to recover it. Mass-loss determinations
depend strongly on the assumed clumping factors and clumping
prescriptions \citep{petrov14,Driessen19}. In addition, empirical analyses often
adopt different velocity laws on either side of the jump, with $\beta$
values switching from $\beta \sim 1$ to $\beta \sim 3$, while most
wind-driving calculations are 1D and do not self-consistently incorporate this
change. Multi-D effects including latitudinal temperature dependence resulting from rotation could also play a role \citep{Gagnier19,Hastings23}.

Systematic uncertainties are likely substantial. Indeed, for individual
objects, different atmosphere codes and modelling strategies often
produce mass-loss rates that differ by up to an order of magnitude (Alkousa et al. 2026). 
Differences in temperature definitions
and reference radii (e.g.\ $T_\star$ versus $T_{\tau=2/3}$) may further
complicate direct comparisons between theoretical and empirical
temperature scales. As a result, a theoretically predicted change in
mass-loss rate occurring over a relatively narrow temperature
interval may be mapped onto a different or shifted range in
empirical studies.

\subsection{Mixed population effects and like-for-like tests}

A fundamentally different issue is that the cool-side B supergiants may not be
the direct descendants of the hot-side O stars. As discussed by
\citet{vink10}, the cool-side objects may represent a distinct
evolutionary population, potentially involving mergers \citep{Menon24}.
In particular, \citet{vink10} discussed two possible interpretations of
the observed $v\sin i$--$T_{\rm eff}$ cliff: bi-stability braking, or two
distinct evolutionary populations.

Recent work indicates systematic differences across the B1 regime,
including (i) a drop in the number of objects on the cool side in
comparison to the hot side \citep{Burgos24} and (ii) indications of a
lower binary incidence on the cool side \citep{Brit25,Patrick25},
consistent with an increasing role for post-interaction products and
mergers \citep{Menon24}. This increasingly favours the second
interpretation for the bulk of the observed $v\sin i$ versus $T_{\rm eff}$
feature \citep{Lennon26}. This does not imply that bi-stability braking
cannot occur in nature, nor that the issue of the terminal age main sequence (TAMS),
especially at higher luminosities, is now resolved \citep{VO25}.
However, it does imply that population-based comparisons are unlikely to be
like-for-like, which would undermine their ability to either prove
or disprove bi-stability jump physics directly.

For instance, if a substantial fraction of objects on the cool side of the
jump are Case B merger products, they may
be underluminous for their mass \citep{Justham14}. Such differences in
$L/M$ (and hence $\Gamma_{\rm e}$) would be expected to affect the wind
properties and inferred mass-loss rates relative to the predominantly
single-star population on the hot side of the feature.

\subsection{Luminosity scaling and the interpretation of trends}

Finally, some of the claimed observational trends may be driven by
secondary parameters. For example, the continuous decrease of $\dot{M}$
with decreasing $T_{\rm eff}$ reported by \citet{Verhamme24} is based on
a small number of objects at low $T_{\rm eff}$, predominantly at very low luminosities. When empirical mass-loss rates are scaled
using luminosity-independent diagnostics such as the transformed mass-loss
rate, the trend is partially \citep{Bernini24} or even fully reversed (Alkousa et al. 2026). 

More generally, empirical constraints on the cool side of the bi-stability jump remain
sparse at higher luminosities. As a result, current samples may not yet
provide a decisive test of whether a bi-stability jump operates in the canonical
B-supergiant regime at fixed $L$. Ultimately, the cleanest empirical test remains a
controlled experiment using individual objects that undergo temperature
excursions at approximately fixed stellar parameters, such as LBVs \citep{vdek02,groh11}.

\section{Discussion and conclusions}

The main result of this note is that hydrodynamically consistent PoWR
models predict a robust bi-stability jump at $T_\star \simeq 25$\,kK even
at relatively low Eddington parameters ($\Gamma_{\rm e}\simeq 0.2$),
confirming that the bi-stability jump is not restricted to the high-$\Gamma$ LBV regime.
Instead, the jump is consistently associated with the
temperature-driven recombination of iron from Fe\,\textsc{iv} to
Fe\,\textsc{iii}, in agreement with earlier Monte Carlo predictions
\citep{vink99,vink00} and CMFGEN calculations \citep{petrov16}. The
persistence of this behaviour across independent CMF-based and
hydrodynamically consistent approaches strengthens the case that the
Fe\,\textsc{iv}/Fe\,\textsc{iii} bi-stability mechanism is a generic
feature of line-driven winds.

At the same time, population-based empirical studies of OB supergiants
have generally not revealed the predicted bi-stability jump signature in mass-loss
rates. We stress that such empirical tests may be fundamentally limited
if the B supergiants on the cool side of the nominal bi-stability jump do not represent
the direct evolutionary continuation of the hot-side O-star population,
as discussed in the two-population scenario of \citet{vink10}. In fact,
growing evidence points to systematic differences across the B1 regime,
including indications of a lower binary incidence on the cool side
\citep{Brit25,Patrick25}, consistent with an increasing role for
post-interaction products and mergers. In this scenario, the absence of a
clear bi-stability jump signature in heterogeneous samples cannot be regarded as a
decisive falsification of the Fe\,\textsc{iv}/Fe\,\textsc{iii}
bi-stability mechanism.

Differences between current comoving-frame implementations remain to be understood and require dedicated benchmarking efforts. Future progress will likely
require (i) targeted code-comparison studies, (ii) improved empirical wind modelling,
in terms of clumping physics and wind hydrodynamics, and (iii) controlled
observational tests using individual objects that undergo temperature
excursions at approximately fixed stellar parameters, such as LBVs.

\begin{acknowledgements}
We thank the anonymous referee for constructive suggestions that helped strengthen the Letter.
JSV and GNS are supported by the STFC grant ST/Y001338/1. 
AACS is supported by the Deutsche Forschungsgemeinschaft (DFG, German Research Foundation) in the form of an Emmy Noether Research Group – Project-ID 445674056 (SA4064/1-1, PI Sander). AACS further acknowledges support by the Federal Ministry of Research, Technology and Space (BMFTR) and the Baden-Württemberg Ministry of Science as part of the Excellence Strategy of the German Federal and State Governments. This project was co-funded by the European Union (Project 101183150 - OCEANS).
\end{acknowledgements}

\bibliographystyle{aa}
\bibliography{references}





\end{document}